\documentclass[trackchanges,twocolumn,twocolappendix]{aastex701}

\usepackage[british]{babel}
\usepackage{booktabs}
\usepackage{latexsym}
\usepackage{amssymb}
\usepackage{amsmath}
\usepackage{graphicx}
\usepackage{tabularx}
\usepackage{cancel}
\usepackage{hyperref}
\usepackage{colortbl}
\usepackage{aas_macros}
\usepackage{multirow}

\usepackage{xcolor}

\usepackage[capitalise]{cleveref}
\usepackage[utf8]{inputenc}
\usepackage[normalem]{ulem}

\newcommand{\Salamanca}{\affiliation{Departamento de F\'isica Fundamental, Universidad de Salamanca, \\Plaza de la Merced, s/n, E-37008 Salamanca, Spain}}
\newcommand{\SalamancaIUFFyM}{\affiliation{Insituto Universitario de F\'isica Fundamental y Matem\'aticas (IUFFyM),\\Universidad de Salamanca, Plaza de la Merced, s/n, E-37008 Salamanca, Spain}}

\begin{document}

\title{Timing Gravity with Pulsars in the Strong Field}

\author{Riccardo~Della~Monica}
\email{rdellamonica@tecnico.ulisboa.pt}
\affiliation{CENTRA, Departamento de Física, Instituto Superior Técnico – IST\\
Universidade de Lisboa – UL, Avenida Rovisco Pais 1, 1049-001 Lisboa, Portugal}

\author{Ivan~De~Martino}
\email{ivan.demartino@usal.es}
\Salamanca
\SalamancaIUFFyM

\begin{abstract}
  We propose a novel approach for the timing of pulsars orbiting a supermassive black hole, which implements the fully relativistic calculations of the photon travel time into a robust timing model. We generate realistic mock catalogues of pulsar times-of-arrival for several putative pulsars on tight orbits around the Galactic Center supermassive black hole, Sagittarius A* (Sgr A*). Then, we perform a proof-of-concept sensitivity analysis to forecast the accuracy that future observational facilities, like the Squared Kilometer Array, will achieve in the characterization of the parameters of our timing model. Our analysis shows how the observation of pulsars at the Galactic Center will open an incredibly promising avenue for the characterization of the physical properties of Sgr A*, which can improve by at least three orders of magnitude the current constraints on the black hole's mass achieved with the S-stars and event-horizon scale observations.
\end{abstract}


\section{Introduction}

The discovery and precise timing of pulsars orbiting a supermassive black hole (SMBH), such as Sagittarius A* (Sgr A*) at the Galactic Center, represents a frontier for testing General Relativity (GR) in the strong-field regime \citep{Stairs2003, Lorimer2008, Will2014}. Pulsars are among the most stable natural clocks, with intrinsic rotational period variations as low as one part in $10^{15}$ per pulse period \citep{Becker2018}. While pulsar timing in non-relativistic binaries has provided valuable tests of GR \citep{Taylor1994, Kramer2006a}, with the recent double-pulsar PSR J0737–3039A/B discovery \citep{Kramer2021, Hu2022} a gold-standard for modern tests of gravity, such systems remain in a comparatively weak-field regime. A pulsar closely orbiting Sgr A*, with periods ranging from a few days to decades, would probe a much stronger gravitational field, potentially allowing precise measurements of the black hole mass, spin, and even higher-order moments \citep{Liu2012, Zhang2017a, Hu202}. Despite extensive radio surveys \citep{Eatough2013a,Torne2023, Lower2024}, only a handful of pulsars have been detected within 15 arcminutes of Sgr A* \citep{Johnston2006,Deneva2009, Kennea2013}. This is attributed to interstellar scattering effects that broaden pulses, reducing detectability at low frequencies \citep{Johnston2006, Deneva2009, Bates2011}. Forthcoming telescopes will benefit from a substantial increase in the collection areas and, among them, the Squared Kilometer Array (SKA) \citep{Eatough2015} has among its major scientific goals the discovery and timing observations of radio pulsars in the Galactic Center \citep{Keane2015}, potentially achieving an accuracy on the measuremeent of pulsar times-of-arrival (TOAs) of the order of $\sigma_\textrm{TOA} \simeq 100$ $\mu$s \citep{Liu2012}.

In anticipation of these discoveries, we have introduced a novel numerical methodology to compute fully relativistic pulsar timing delays in spherically symmetric spacetimes \citep{DellaMonica2023d}. Previous pulsar timing forecast analyses for pulsars at the Galactic Center have relied on post-Newtonian (PN) approximations \citep{Liu2012, Hu:2023ubk, Hu:2026zcb}, which become increasingly inaccurate in the strong gravitational field of an SMBH \citep{DellaMonica2025}. Our method, on the other hand, is based on fully-relativistic geodesic calculations for both photon propagation and pulsar motion, allowing precise predictions of TOAs. Since the integration of the relevant equations is undertaken numerically in our framework, it can be applied generically for any spherically-symmetric spacetime metric describing black holes (or mimickers) in General Relativity or even in alternative theories of gravity. In this Letter we perform a sensitivity analysis of the possible pulsar observations in the Galactic Center that SKA will deliver in the next decade. We demonstrate that with such future observations one can achieve a characterization of the physical properties of Sgr A*, that enables an estimation of the mass that promises to be at least three orders of magnitude more precise than the one currently achievable with the S-stars and the event-horizon scales observations in the Galactic Center \citep{DeLaurentis2023}.

\emph{Mock data and posterior analysis}. --- The pulsar timing analysis relies on precise calculations of pulse TOAs at a distant observer's location. Our aim is to carry out timing analysis of putative pulsars existing in the Galactic Center, on tight orbits around Sgr A*. For this reason, we consider the observer at a distance of $r_\textrm{o}=8$ kpc, we set the mass of the central object to $M = 4.261\times10^6M_\odot$ \citep{GravityCollaboration2020c}, and we introduce a series of pulsar toy models, that we have previously analyzed \citep{DellaMonica2023d}, whose orbital parameters are listed for completeness in Table \ref{tab:pulsars_toy_models}. The first toy model has orbital parameters resembling those of the S2 star in the Galactic Center \citep{DeLaurentis2023}, all the other models are intended to probe increasingly strong gravitational regimes as quantified by the relativistic parameter $\Gamma\equiv r_g/r_p = GM/ac^2(1-e)$ and the orbital velocity parameter\footnote{For comparison, the orbital velocity parameter for the double pulsar system J0737–3039A/B \citep{Kramer2021, Hu2022} is estimated to be $\beta_\textrm{O}^2 = 4.3\times10^{-6}$.} $\beta_\textrm{O}^2 = (2\pi GM/c^3T)^{2/3}$ \citep{Damour1992}.

In the present proof-of-concept analysis, the pulsar is modeled as a massive test particle moving on a geodesic of the background spacetime. This approximation is justified by the extreme mass ratio of the system. Assuming a canonical pulsar mass of $m \approx 1.4\,M_\odot$, yields a mass ratio $q\equiv{m}/{M}\simeq3.3\times10^{-7}$. The dissipative effects associated with gravitational-wave emission are therefore expected to be extremely small over the timescale of the observations considered in this Letter. For instance, using the Peters–Mathews approximation \citep{Peters:1964zz}, the shortest inspiral timescale among the orbital configurations considered (Toy 3) is of the order of $t_{\rm GW}\sim 3\times10^{8},{\rm yr}$. Since we consider a mock timing campaign that span only two years, the corresponding secular orbital evolution satisfies $\Delta a/a\lesssim10^{-8}$ even in the most relativistic case considered, and can therefore be safely neglected. Similarly, our model does not include conservative finite-mass effects, such as the gravitational self-force \citep{Poisson:2011nh, Barack:2018yvs}, nor spin-curvature couplings described by the Mathisson–Papapetrou–Dixon equations \citep{Papapetrou:1951pa, Dixon:1970zza}. These corrections are expected to scale with the small mass ratio $q$ and are therefore subleading with respect to the leading-order orbital dynamics.

With these caveats in mind, we can assume that both the trajectory of the pulsar and of the photons emitted by it can be described as geodesics in the gravitational field generated by the central object. In this work, we consider that the space-time surrounding the SMBH in the Galactic Center can be described by the general relativistic Schwarzschild solution which, in harmonic coordinates (usually employed for pulsar timing models) and geometrized units ($G=c=1$), takes the form
\begin{equation}
  ds^2 = -\left(\frac{r-M}{r+M}\right)dt^2 + \left(\frac{r+M}{r-M}\right)dr^2+(r+M)^2d\Omega^2,
  \label{eq:metric}
\end{equation}
where $d\Omega^2=d\theta^2+\sin^2\theta d\phi^2$ represents the solid angle element, $t$ is the coordinate time measured by an infinitely-distant stationary observer and $r$ is the harmonic radius, related to the usual aerial radius by $r_S=r+M$, being $M$ the mass of the central object. Freely falling test particles in the space-time described by the line element in Eq. \eqref{eq:metric} follow geodesic trajectories. For the motion of the pulsar we directly solve the geodesic equations numerically, using our open-source tool \href{https://github.com/rdellamonica/pygro}{\texttt{PyGRO}} \citep{PyGRO2025}. In particular, we adopt a mixed analytical-numerical methodology to fix initial conditions for the time-like geodesic based on its orbital parameters, equivalent to the usual Keplerian celestial mechanics parametrization of elliptic orbits, and then integrate the relativistic equations of motion numerically. These parameters are the coordinate time of pericenter passage, $t_P$, the semi-major axis, $a$, the eccentricity, $e$, and three angular parameters, $i$, $\Omega$ and $\omega$, the inclination, the longitude of the ascending node and the argument of the pericenter, respectively, which uniquely identify an orbital plane in space-time.
For the photons emitted by the pulsar and received by a distant observer, on the other hand, we resort to a methodology valid for spherically-symmetric space-times introduced in \citet{DellaMonica2023d} and further developed in \citet{DellaMonica2025}. The key challenge in relativistic pulsar timing is solving the emitter-observer problem: \emph{i.e.} finding the correct null geodesic connecting the pulsar at emission to the observer at reception. This requires solving the integral equation
\begin{equation}
  \phi_{\textrm{o}}-\phi_{\textrm{e}} =
  \int_{r_{\textrm{e}},\gamma}^{r_{\textrm{o}}}\frac{b}{r\sqrt{\Upsilon(r)}}dr.
  \label{eq:phi_integral}
\end{equation}
for the unknown impact parameter of the photon $b$. Here, $\Upsilon(r)$ is a function of the radial coordinate solely depending on the components of the metric tensor and depending parametrically on $b$ \citep{DellaMonica2023d}, $\gamma$ represents the photon path and quantities with a subscript $e$ ($o$) represent coordinates of emission (observation) of the photon. Once the photon trajectory is determined, the propagation time is obtained from
\begin{equation}
  \Delta t \equiv t_{\textrm{o}}-t_{\textrm{e}} = \int_{r_{\textrm{e}},\gamma}^{r_{\textrm{o}}} \frac{r(r+M)}{(r-M)\sqrt{\Upsilon(r)}}dr\,.
  \label{eq:relativistic_propagation_time}
\end{equation}
We solve for these quantities numerically. In particular, the emission coordinates ($t_\textrm{e}$, $r_\textrm{e}$, $\phi_\textrm{e}$) are obtained from the numerically integrated pulsar geodesic, and are used as input to solve the emitter–observer problem for the photon trajectory. Eq.~\eqref{eq:phi_integral} determines the photon impact parameter $b$ that connects the emission point on the pulsar trajectory to the observer, while Eq.~\eqref{eq:relativistic_propagation_time} yields the corresponding propagation time along that null geodesic. A full derivation of the algorithm and implementation details are provided in \citet{DellaMonica2025}.

Our procedure yields a fully-relativistic timing function which maps proper times of emissions $\tau_i$ into TOAs at the observer's location
\begin{equation}
  \textrm{TOA}_i = \textrm{TOA}(\tau_i),
  \label{eq:timing_function}
\end{equation}
in which the full non-linearity of the combination of relativistic effects on the trajectory and the photon is taken into account.

\begin{table}
  \footnotesize
  \centering
  \setlength{\tabcolsep}{5.2pt}
  \renewcommand{\arraystretch}{1.5}
  \caption{Orbital parameters for the toy models used in our analysis (we also consider for all the models an inclination of $i = \pi/3$ and $\omega=\Omega=0$). The four objects probe increasingly strong gravitational regimes, as quantified by the relativistic parameter $\Gamma\equiv r_g/r_p = GM/ac^2(1-e)$ and the orbital velocity parameter $\beta_O^2 = (2\pi GM/c^3T)^{2/3}$, reported in the last two columns.}
  \label{tab:pulsars_toy_models}
  \begin{tabular}{lccccc}
    \hline
    \textbf{Model} & $a$ (AU) &  $e$   & $T$  &   $\Gamma$ ($10^{-4}$)  & $\beta_O^2$ ($10^{-4}$) \\ \hline
    {Toy 0} & 1025  & 0.88   & 16 yr  & 3 & 0.41\\
    {Toy 1} & 175.4  & 0.800   & 1.162 yr & 11 & 2.3 \\
    {Toy 2} & 80.0  & 0.800  & 126 days & 25 & 5.3\\
    {Toy 3} & 43.8  & 0.800  & 52.9 days & 45 & 9.4\\ \hline
  \end{tabular}
\end{table}

We use the the timing model described above both to generate catalogs of mock observations for the toy models in Table~\ref{tab:pulsars_toy_models}, mirroring the sensitivity of a future SKA-like observatory, and to later fit the timing model to those catalogs. In particular, for each toy model we consider a total number $N_\textrm{data}=120$ of TOAs acquired in four observation runs with a duration of one week each, taken 6 months apart from each other, thus covering a total time-span of 2 years\footnote{The adopted observing schedule is intended only as a representative example of a realistic monitoring campaign; a detailed optimization of observational strategies and their impact on parameter recovery is deferred to future work.} Additionally, we assume intrinsic pulsar period of $P=2$ s and a period spin-down rate of $\dot{P} = 10^{-15}$ s s$^{-1}$. Finally, without loss of generality, in each mock catalogue we assume that the first pulse arrives at the Earth-based observatory exactly at year 2025.0. On each TOA, we then include the instrumental uncertainty of SKA by adding a Gaussian noise with $\sigma_\textrm{TOA} = 100\,\mu$s \citep{Liu2012}. The result is a list of $N_\textrm{data}$ numbers representing the TOAs in MJD units with the corresponding uncertainty, as it is customary for pulsar timing data.

\begin{table*}[!ht]
  \footnotesize
  \centering
  \setlength{\tabcolsep}{18pt}
  \renewcommand{\arraystretch}{1.4}
  \begin{tabular}{lcccc}
    \hline
    \textbf{Parameter} (unit) & \textbf{ Posterior} & \textbf{ Precision} (\%) & \textbf{ Posterior}  & \textbf{ Precision} (\%) \\
    \hline
    & \multicolumn{2}{c}{\textbf{Toy 0} (S2-like)}  & \multicolumn{2}{c}{\textbf{Toy 1}}
    \\ \hline
    $M$ ($M_\odot$) & $4261000.3(2)$ & $5.2 \times 10^{-5}$ & $4260999.965(87)$ & $2.1 \times 10^{-6}$ \\
    $a$ (AU) & $1025.0004(31)$ & $0.0003$ & $175.3999994(14)$ & $8.2 \times 10^{-7}$ \\
    $e$ & $0.88000004(35)$ & $3.9 \times 10^{-5}$ & $0.800000001(20)$ & $2.5 \times 10^{-7}$ \\
    $P$ (s) & $1.9999999999(11)$ & $5.3 \times 10^{-8}$ & $2.00000000006(13)$ & $6.5 \times 10^{-9}$ \\
    $\dot{P}$ ($10^{-15}$ s) & $0.99(11)$ & $11$ & $0.9981(43)$ & $0.43$ \\ \hline
    & \multicolumn{2}{c}{\textbf{Toy 2}} & \multicolumn{2}{c}{\textbf{Toy 3}} \\ \hline
    $M$ ($M_\odot$) & $4260999.939(70)$ & $1.6 \times 10^{-6}$ & $4261000.05(47)$ & $1.1 \times 10^{-6}$ \\
    $a$ (AU) & $79.99999959(47)$ & $5.8 \times 10^{-7}$ & $43.80000018(16)$ & $3.7 \times 10^{-7}$ \\
    $e$ & $0.800000003(25)$ & $3.1 \times 10^{-7}$ & $0.8000000045(43)$ & $5.3 \times 10^{-7}$ \\
    $P$ (s) & $2.000000000057(57)$ & $2.9 \times 10^{-9}$ & $1.999999999977(31)$ & $1.5 \times 10^{-9}$ \\
    $\dot{P}$ ($10^{-15}$ s) & $0.9985(14)$ & $0.14$ & $1.00049(53)$ & $0.053$ \\ \hline
  \end{tabular}
  \caption{Best fitting values and (within parenthesis) $1\sigma$ uncertainties resulting from our MCMC analysis on the five-parameters timing model for the pulsar toy models in Table \ref{tab:pulsars_toy_models}. We also report, for each toy model, the percentage precision reached on the parameters (given by the ratio of the parameter uncertainty to its best-fitting value) achieved during the two years of observations .}
  \label{tab:pulsar_posterior}
\end{table*}

After generating the mock catalogue, we address the problem of performing a posterior analysis on the synthetic dataset, to assess the accuracy and the precision by which the intrinsic and orbital parameters of the pulsar and the mass of the central object can be recovered. In order to do so, one has to define an inverse timing function
\begin{equation}
  \tau_i = \tau(\textrm{TOA}_i),
  \label{eq:inverse_timing_function}
\end{equation}
which, from the measured $\textrm{TOA}_i$ allows to compute the proper time of emission of the observed photon $\tau_i$ and thus compute phase-connected timing residuals on which to perform statistics. The problem of how to invert the timing function in Eq. \eqref{eq:timing_function} in a relativistic scenario, in comparison with the much simpler case of adopting a post-Newtonian approximation, has been tackled in \citep{DellaMonica2025}. Here we only give a brief overview of the numerical methodology. The inverse timing formula depends parametrically on the following parameters
\begin{equation}
  \vec{\theta} = \left(P,\,\dot{P},\,\tau_P,\,\Phi_0,\,M,\,t_p,\,a,\,e,\,i,\,\omega,\,\Omega\right).
\end{equation}
Along with the the Keplerian parametrization of the initial conditions for the pulsar's trajectory and the intrinsic pulsar parameters (period and spin-down rate), we also consider a temporal offset parameter $\tau_P$ corresponding to the instant of proper time at which the pulsar has rotational period $P$ and a zero-point offset $\Phi_0$ representing the pulsar phase for the first received photon. The distance of the observatory from the SMBH, $r_\textrm{o}$ only contributes to a constant shift of the photon travel times (as long as $r_\textrm{o}\gg2M$), which can be included in the phase shift $\Phi_0$. Therefore, the distance will not be considered as a free parameter of our model. To obtain the inverse timing function from these parameters, first, we integrate the geodesic equations for the pulsar's orbit numerically. Then we sample a dense set of a total of $N_\textrm{dense}$ emission times on the integrated trajectory. These sampled times are a subset of the total number of pulses ($N_\textrm{pulses}$) emitted over the observation period. For instance, a pulsar with a 2-second pulse period can emit approximately $N_\textrm{pulses}\sim1.84 \times 10^7$ pulses per orbital period for an orbit similar to the Toy 1 model in Table \ref{tab:pulsars_toy_models}. To each sampled point we apply the method mentioned above to solve the emitter-observer problem and to calculate the relativistic propagation time. The total TOA at the observer is then computed as the sum of the coordinate emission time and the relativistic travel time. This process yields a sampling of the direct timing formula, $\textrm{TOA}(\tau)$, which we thus reconstruct at $N_\textrm{dense}$ discrete points. We can use these samples to interpolate the function 
and invert it numerically, thus yielding a reconstruction of the inverse timing formula in Eq. \eqref{eq:inverse_timing_function}. A crucial parameter of this methodology is the number of sample points $N_\textrm{dense}$, which must satisfy the inequality $N_\textrm{data} \leq N_\textrm{dense} \leq N_\textrm{pulses}$, where $N_\textrm{data}$ is the number of actual TOA measurements. While using $N_\textrm{dense} = N_\textrm{pulses}$ is the choice that would provide the maximum accuracy, \emph{i.e.} a 1:1 reconstruction of the timing function, in practical scenarios integrating this many photon paths is not feasible. In \citet{DellaMonica2025} we have devised a convergence procedure showing that for our toy models a choice of $N_\textrm{dense}\sim 500 N_\textrm{data}$, provides a good balance between computational cost and precision for the accuracy at which we aim.

Once the inverse timing formula is obtained, we can finally compute the pulse phase for all the photons, as done in usual timing codes,
\begin{align}
  \Phi_i(\tau_i) &= \phi_0 + \frac{\tau_i-\tau_P}{P} - \frac{\dot{P}}{2P^2}(\tau_i-\tau_P)^2,
  \label{eq:pulse_phase_model}
\end{align}
and define the timing residuals as
\begin{equation}
  R_i = \frac{\Phi_i-N_i}{\nu} = (\Phi_i-N_i)P,
  \label{eq:pulsars_residuals}
\end{equation}
corresponding to the time since (or to) the nearest actual pulse emitted by the pulsar. The residuals allow us to define a likelihood function
\begin{equation}
  \log\mathcal{L} = -\frac{1}{2}\sum_{i=1}^{N_\textrm{data}}\left(\frac{R_i}{\sigma_i}\right)^2,
  \label{eq:likelihood_pulsars}
\end{equation}
which we use to perform a Monte Carlo Markov Chain (MCMC) analysis, to fit our timing model to the TOAs mock data. In particular we adopt the affine invariant ensemble sampler implemented in the Python package \texttt{emcee} \citep{ForemanMackey2013}. We adopt uniform large priors on all the parameters of our analysis, which are heuristically based on the qualitative analysis of the precision achieved with SKA that we have presented in \citep{DellaMonica2025}. We run MCMC simulations on all the toy models in Table \ref{tab:pulsars_toy_models} to estimate the accuracy down to which SKA will be able to measure intrinsic pulsar parameters, the black hole mass, and the orbital parameters. As a proof of concept, we only consider a subset of five free parameters, $(M,\,a,\,e,\,P,\,\dot{P})$, fixing the remaining ones to their \emph{true} values.

\begin{figure}[!t]
  \includegraphics[width=\columnwidth]{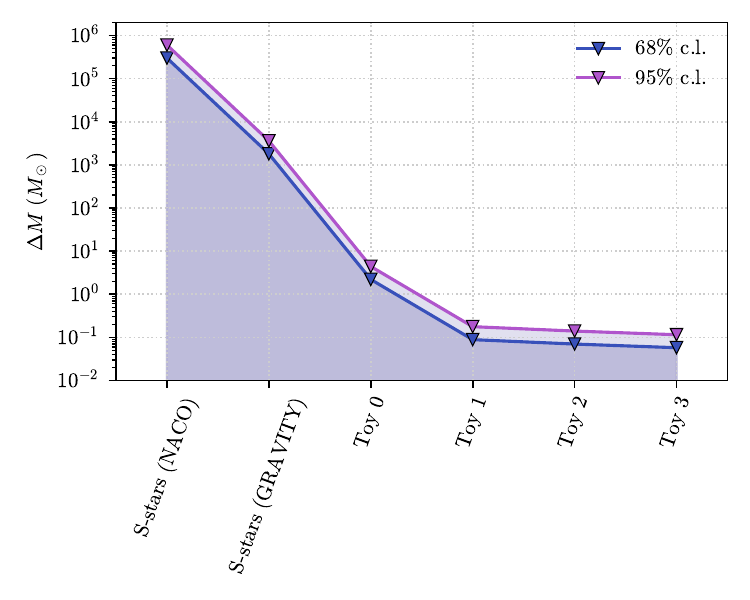}
  \caption{Posterior distributions for the mass of Sgr A* relative to the input (true) values for all toy models. The 68\% (blue line) and 95\% (pink line) median-centered confidence intervals are shown, indicating that the true parameter values consistently lie within the recovered interval across all models. We also display, for comparison, the precision on $M$ achieved from the astrometric and spectroscopic data for the S2 star, with either the NACO imager or the GRAVITY interferometer. The precision on the recovery of the true value of $M$ achieved with pulsar timing is by several orders of magnitude higher than that achieved with S-stars data. Moreover, the precision  improves significantly, by almost two additional orders of magnitude, from Toy 0 to Toy 3, demonstrating that stronger gravitational regimes yield tighter constraints from relativistic pulsars timing analyses.}
  \label{fig:pulsar_precision}
\end{figure}

The full set of results from our posterior analyses on the mock catalogue for the toy models in Table \ref{tab:pulsars_toy_models} is reported in Table \ref{tab:pulsar_posterior}, in which we list the best fit and the resulting precision on the recovery of the input parameters. The posteriors always encompass the true value of parameters within the 68\% confidence interval, showing no biases emerging from our analysis going at increasingly strong gravitational regimes. Strikingly, the precision on the estimation of the mass, along with all the other intrinsic and orbital parameters, improves by several orders of magnitude going from Toy 0 to Toy 3, supporting our conclusion that the stronger the gravitational probe used, the tighter the constraints on the properties of Sgr A*. Focusing on the estimation of the mass of Sgr A*, in Figure \ref{fig:pulsar_precision} we report the amplitude of the median-centered 68\% and 95\% confidence intervals on this parameter for all the toy models considered. For comparison, we report in the same plot the equivalent amplitude of the confidence interval currently achieved with observations of the S-stars with the NACO adaptive-optics imager \citep{Gillessen2017} or the GRAVITY interferometer \citep{GravityCollaboration2024}. The fact that our model Toy 0 has the same orbital parameters as the S2 star (the one with most constraining power in the cluster) is remarkable, as the constraints on the mass of Sgr A* from pulsar timing surpass by almost three orders of magnitude the ones obtained with orbital fitting for these objects. This highlights the revolutionary impact of the successful timing of a pulsar around Sgr A* for the characterization of its physical properties, even on a mildly relativistic orbit. A corner plot with contours depicting the full posterior distribution of our MCMC analyses is shown for completeness in the Appendix.

\emph{Discussion and Conclusions}. ---
The endeavor to test GR in the strong-field regime has historically relied on binary pulsars and the orbital dynamics of stars around SMBHs \citep{Will2014, DeLaurentis2023}. While the timing of binary pulsars has yielded some of the most stringent tests of GR to date \citep{Taylor1994, Kramer2006a, Kramer2021, Hu2022}, these systems remain within the relatively weak gravitational regime. In contrast, stars like S2 in the Galactic Center \citep{GravityCollaboration2024} have probed stronger fields, but their use as test particles is limited by astrometric precision and the comparatively low cadence of observational data \citep{DeLaurentis2023}. Pulsars in close orbits around the SMBH at the Galactic Center --- though still undetected ---represent a uniquely powerful probe, which combines the compactness and temporal regularity of pulsars with the extreme gravitational regime in the vicinity of Sgr A* \citep{Liu2012, DellaMonica2023d, DellaMonica2025}. In this work, we have presented a novel numerical framework for pulsar timing in the strong-field regime of a spherically symmetric spacetime, implementing fully relativistic geodesic calculations for both the motion of the pulsar and the propagation of emitted photons. This methodology goes beyond standard post-Newtonian approximations \citep{Damour1986, Hobbs2006}, which are known to break down in the vicinity of compact massive objects \citep{Carleo2025, DellaMonica2025}, and enables the computation of precise TOAs predictions directly from first principles. We generated mock pulsar TOA datasets for a suite of increasingly relativistic orbits, modelled to mirror the prospective capabilities of the SKA. The results of our analysis are remarkable: even for orbits with parameters comparable to the well-studied S2 star \citep{DeLaurentis2023} (corresponding our Toy~0 model), the inferred precision on the mass of Sgr A* improves by nearly three orders of magnitude over current constraints from S-star astrometry, both with NACO and GRAVITY \citep{GravityCollaboration2024}. As the orbital period decreases and the relativistic parameter $\Gamma$ increases, the precision on all recovered parameters (including the mass of the SMBH, orbital elements, and intrinsic pulsar spin parameters) is predicted to improve even further, by two additional orders of magnitude, reaching unprecedented levels, with a subpart per million accuracy in several cases, which is broadly consistent with previous forecast analyses \citep{Liu2012, Hu:2026zcb} within the post-Newtonian approximation.

We emphasize that the uncertainties that we report here represent statistical sensitivities within the adopted timing model. A complete analysis of future pulsar observations in the Galactic Center will require the inclusion of additional physical effects beyond the geodesic approximation, including conservative self-force corrections \citep{Peters:1964zz, Barack:2018yvs} and spin-curvature couplings \citep{Papapetrou:1951pa, Dixon:1970zza}. While these effects are not expected to alter the qualitative conclusions of the present work, they may ultimately contribute to the systematic-error budget once parameter estimation accuracies approach the extreme mass-ratio scale of the system.

A further limitation of the present analysis is the assumption of spherical symmetry. This assumption allows us to isolate and validate the strong-field timing methodology in a minimal setting. The spacetime of Sgr A* is expected to be more accurately described by the Kerr solution, in which case frame-dragging effects and the associated quadrupolar deformation of the spacetime would introduce additional timing signatures beyond those considered here \citep{Liu2012, Wex2014, Izmailov2019}. These effects not only can constitute sources of systematic error but also represent valuable observables, as they enable direct tests of the no-hair theorem. Extending the present framework to axisymmetric spacetimes (as in \citet{BenSalem2022}) is part of ongoing work and represents a natural next step for the numerical approach adopted here.

Another source of potential systematic uncertainty arises from environmental effects in the Galactic Center. Nearby stars and compact remnants \citep{Merritt:2009ex}, third-body perturbations  \citep{Stephan:2016kwj}, extended mass distributions and dark-matter structures \citep{Sadeghian:2013laa, Hu:2023ubk}, or gaseous components \citep{Psaltis2012} may introduce timing signatures that partially correlate with relativistic effects generated by the SMBH itself \citep{Eatough2015}. Assessing the detectability and degeneracy of these perturbations is beyond the scope of the present Letter. Nevertheless, the numerical methodology developed here is not restricted to vacuum spacetimes and can naturally accommodate additional dynamical forces acting on the pulsar.

Our findings underscore the transformative potential of pulsar timing in the Galactic Center. Not only does it offer the most accurate measurement of the mass of Sgr A* to date \citep{DeLaurentis2023}, but it also provides a pathway to precisely test deviations from general relativistic metrics, constrain alternative theories of gravity \citep{DellaMonica2023d}, and detect the imprint of higher-order multipole moments of the central object \citep{Vigeland2010}. Remarkably, the methodology we developed remains fully valid for arbitrary spherically symmetric metrics, and can be naturally extended to rotating spacetimes, crucial for studying frame-dragging and the no-hair theorem.

\clearpage

\begin{acknowledgments}
RDM acknowledges financial support provided by FCT – Fundação para a Ciência e a Tecnologia, I.P., through the ERC-Portugal program Project ``GravNewFields''. RDM also thanks the Fundação para a Ciência e Tecnologia (FCT), Portugal, for the financial support to the Center for Astrophysics and Gravitation (CENTRA/IST/ULisboa) through grant No.~\href{https://doi.org/10.54499/UID/PRR/00099/2025}{UID/PRR/00099/2025} and grant No.~\href{https://doi.org/10.54499/UID/00099/2025}{UID/00099/2025}. IDM acknowledges support from the grant PID2024-158938NB-I00 funded by MCIN/AEI/10.13039/501100011033 and by ``ERDF A way of making Europe''. IDM also acknowledges support from the grant SA097P24 funded by Junta de Castilla y Le\'on and by "ERDF A way of making Europe". We finally thank high-performance computing resources of the Castilla y Le\'on Supercomputing Center (SCAYLE), \url{www.scayle.es} for providing supercomputing facilities.
\end{acknowledgments}

\facilities{Castilla y Le\'on Supercomputing Center (SCAYLE), \url{www.scayle.es}}

\software{PyGRO \citep{PyGRO2025}}

\bibliography{biblio}{}
\bibliographystyle{aasjournalv7}

\appendix

\section{Posterior distributions}
\begin{figure}[!b]
  \includegraphics[width=\columnwidth]{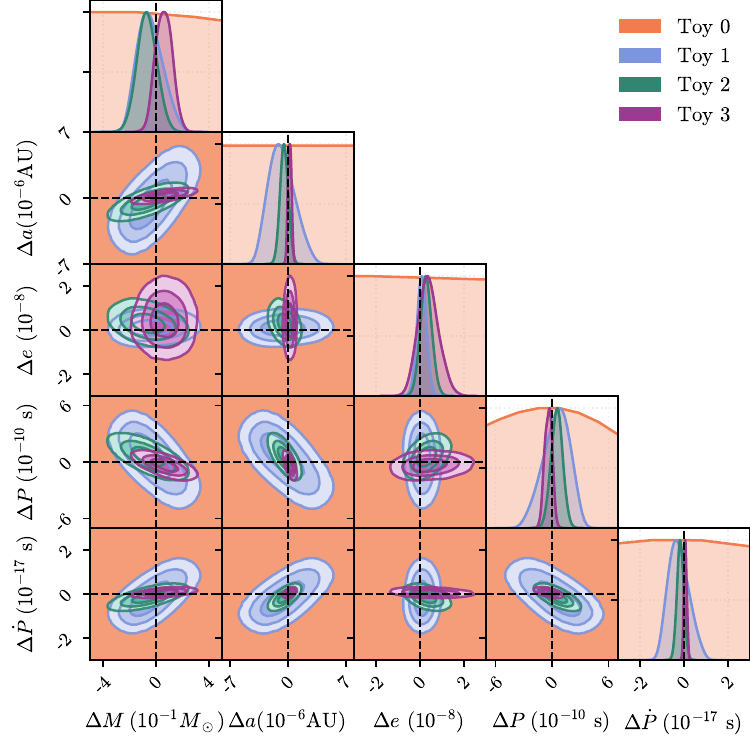}
  \caption{Full five-dimensional posterior distribution from our MCMC analyses on the mock catalogues for all the pulsar models considered in Table \ref{tab:pulsars_toy_models}. The black dashed lines report the \emph{true} values of the parameters, used as input for the generation of the mock catalogue. These are correctly recovered within 1$\sigma$ by our methodology, thus showing that not only we reach incredibly precise, but also accurate results. We display all posteriors relative to their true values, thereby all contours are centered on 0. The Toy 0 posteriors (in orange) exceed the limits of the plot showing the significant improvement in the parameters estimation provided by pulsars on tighter orbits with respect to S2.}
  \label{fig:pulsar_posteriors}
\end{figure}
We report in Figure \ref{fig:pulsar_posteriors} the corner plot of the full posterior distribution from our MCMC analysis for all the toy models considered, relative to their medians. The black dashed lines correspond to the true values for each parameter, used for the generation of the mock catalogues, showing that the \emph{true} values are recovered within $1\sigma$ accuracy with no significant bias. For Toy 0, the posteriors (in orange) shown in the plots exceed the considered ranges for all the parameters. This shows the significant improvement in the estimation of the parameters of the timing model provided by pulsars on tighter orbits with respect to the S2 star. If the limits were instead adjusted to include the full range of the posterior for Toy 0, the posteriors for all other toy models considered would reduce to a point. The contours became increasingly tighter as the gravitational regime probed by the pulsar gets stronger. The only parameter that shows no significant improvement from Toy 1 to Toy 3 is the orbital eccentricity $e$, which results to be marginally less constrained for Toy 3, compared to Toy 2. We argue this can be the consequence of how secular effects emerging from mis-estimations of the eccentricity evolve over time (see the discussion in Section IV.A of \citet{DellaMonica2025}) in relation to the observational strategy that we have chosen to adopt (one-week-long observation runs, 6 months apart), which results in a different sampling of the orbital phase for the different toy models.

\end{document}